\documentclass[prl,showpacs,twocolumn,superscriptaddress]{revtex4}
\usepackage{graphicx,amssymb,amsmath,color}
\begin{document}
\title{Temperature dependent conductances of deformable molecular devices}

\author{Bal\'azs D\'ora}
\email{dora@pks.mpg.de}
\affiliation{Max-Planck-Institut f\"ur Physik komplexer Systeme, N\"othnitzer Str. 38, 01187 Dresden, Germany}
\author{Andr\'as Halbritter}
\affiliation{Department of Physics, Budapest University of Technology and Economics and Condensed Matter Research Group of the
Hungarian Academy of Sciences, 1111 Budapest, Budafoki \'ut 8, Hungary}

\date{\today}

\begin{abstract}

Transport through a molecular device coupled to a vibrational mode is studied. By mapping it to
the Yu-Anderson model in the large contact broadening limit, the zero bias electric and heat conductances are evaluated
non-perturbatively. These exhibit a
step from their $T=0$ value to half
of its value as $T$ increases due to the opening of the inelastic scattering channel. 
The spectral function exhibits the Franck-Condon suppressed multiphonon steps.
The Wiedemann-Franz law is satisfied at low
and high temperatures, but is violated in
between.
Relations to experiments are discussed.


\end{abstract}

\pacs{73.23.-b,73.63.-b,71.38.-k,85.65.+h}

\maketitle

\textit{Motivation.}
Molecular electronic devices are based on electron transport through individual molecules, and have been proposed for many 
years as possible candidates for future
nanoelectronic circuit elements \cite{reed,nitzan}. They  triggered intense research due to the 
fundamental challenge of quantum transport in nanoscale systems and by the possibility of tailoring molecular scale 
structures\cite{cuniberti,galperin}.
One of their most appealing features is the importance of local Coulomb interaction between quasiparticles and scattering off local 
vibrational modes, which are more pronounced due to the less effective screening mechanisms at small length scales.

A wide range of characterization methods have 
been developed for collecting information about the microscopic properties of the molecular junctions and probing its excited states,
including conductance histogram techniques  \cite{agrait}; Coulomb blockade \cite{park,pasupathy}, 
conductance fluctuation \cite{smit}, shot noise \cite{djukic} or superconducting subgap 
structure \cite{makk} measurements, operating mainly at low temperatures.
 An especially important information about molecular 
junctions is the study of vibrational modes with point-contact or inelastic electron tunneling 
spectroscopy \cite{smit,tal}, where the low-temperature zero-bias conductance of the device is 
perturbed by the excitation of molecular vibrations with an appropriate bias voltage. Indeed, 
experiments on single molecular devices including fullerenes\cite{park,pasupathy}, carbon 
nanotubes\cite{leroy}, benzene\cite{kiguchi} and simple molecules (H$_2$, CO, H$_2$O) connecting 
metallic electrodes\cite{smit,tal}, have revealed the influence of vibrational degrees of freedom 
supplemented by theoretical simulations of the vibrational 
modes\cite{flensberg,koch,paulsson,vega,eggergog,cornaglia1,chen,galperin}.

Whereas the above low-temperature techniques provide a large amount of information about the ground 
state properties of the device and small perturbations due to the excitations of the inelastic degrees 
of freedom, any room temperature application requires the knowledge of the temperature sensitivity of the
device\cite{nitzan} and the impact of the strongly enhanced inelastic scattering 
processes should be considered\cite{cuniberti}. Therefore, in this work, we focus on the temperature dependent transport 
through molecular devices by studying  the electrical and thermal conductance of a model system 
with non-perturbative analysis, valid for arbitrary 
electron-vibron couplings.

\textit{The model and its mapping.} To model a molecular transport junction, we consider the molecular
Hamiltonian as
\begin{equation}
H_{mol}=g_{v}Qd^+d+\frac{P^2}{2m}+\frac{m\omega_0^2}{2}Q^2,
\label{hmol}
\end{equation}
describing a single electron level coupled to an Einstein phonon. The level is tunnel coupled to leads as
\begin{equation}
H_0=\sum_{k}\varepsilon(k)c^+_{k}c_{k}+ \sum_k V_k\left(c_k^+d+d^+c_k\right)+\epsilon_0 d^+d,
\end{equation}
where $V_k$ is the hybridization parameter. The contact broadening is given by
$\Delta=\pi\rho_0\langle V_k^2\rangle_{FS}$.
Let us consider symmetric contacts with non-interacting quasiparticles, in which case it suffices to consider a single
effective contact\cite{giuliano} with energy dispersion $\varepsilon(k)=v_Fk$ and density of states (DOS) $\rho_0=1/2\pi v_F$.
First, we diagonalize $H_0$, and express the molecular Hamiltonian in terms of its eigenfunctions.
This is achieved by introducing\cite{mahan}
\begin{gather}
d=\sum_k\nu_k a_k,\hspace*{5mm} c_k=\sum_{k'}\eta_{k,k'}a_{k'},
\label{trafo}
\end{gather}
and
\begin{gather}
\nu_k^2=\frac{2v_F}{L}\frac{ \Delta}{(\varepsilon_k-\epsilon_0)^2+\Delta^2},
\end{gather}
where $L$ is the size of the sample. The explicit form of $\eta_{k,k'}$ is irrelevant for our discussion\cite{mahan}.
From now on, we restrict our attention to the cases when the broadening by the contacts is such\cite{paulsson} that the DOS of the
contact and the device are slowly varying on the scale of the phonon energy $\omega_0$ and temperature $T$.
This is realized in the case $\Delta\gg (T,\omega_0)$ and $\Delta>|\epsilon_0|$, when the junction is represented by a
resonant level
model, thus
$\nu_{k}\sqrt{L}\approx \bar
\nu=\sqrt{2v_F\Delta/(\epsilon_0^2+\Delta^2)}$ is $k$
independent, which are the basic conditions for the mapping to hold.
These are often satisfied under realistic conditions\cite{paulsson,vega}.
Other cases were studied in Ref. \onlinecite{cornaglia1,chen,koch}.
In terms of the transformation in Eq. \eqref{trafo}, the total Hamiltonian $H_0+H_{mol}$ is rewritten as
\begin{equation}
H=\sum_{k}\varepsilon(k)a^+_{k}a_{k}+\bar{g} Q\Psi^+({0})\Psi({0})+\frac{P^2}{2m}+\frac
{m\omega_0^2}{2}Q^2,
\label{hamilton}
\end{equation}
where $\bar g={\bar \nu}^2 g_v$ and $\Psi({x=0})=\sum_{k}a_k/\sqrt{L}$. Eq. \eqref{hamilton} is known as the Yu-Anderson
model\cite{yuanderson}, describing conduction
electrons
interacting with a local bosonic mode. The model is solved by
bosonizing the fermionic field\cite{delft,doraphonon} as
$\Psi(x)=\exp[i\sqrt{4\pi}\Phi(x)]/\sqrt{2\pi\alpha}$, and
the
resulting effective model of one dimensional coupled harmonic oscillators\cite{weiss} reads
as
\begin{equation}
H=v_F\int\limits_{-\infty}^{\infty}\textmd{d}x
\left[\partial_x\Phi(x)\right]^2+\frac{g}{\sqrt\pi}Q\partial_x\Phi(0)+\frac{P^2}{2m}+\frac{m\omega_0^2}{2}Q^2,
\label{hamboson}
\end{equation}
where $g$ is the phase shift caused by $\bar g$, $\Phi(x)$ stems from the bosonic representation of
the
fermion field.
The vibrational mode softens, and the damped frequencies are given by \cite{doraphonon}
\begin{equation}
\omega_{p\pm}=-i\Gamma\pm\sqrt{\omega_0^2(1-\Gamma/\Gamma_2)-\Gamma^2},
\end{equation}
where $\Gamma_2=\pi\omega_0^2/4W\ll\omega_0\ll W$, $W$ is the bandwidth of the conduction electrons, and
$\Gamma=\pi(g\rho_0)^2/2m$ for small $g$, and the model becomes unstable for $\Gamma>\Gamma_2$.
The explicit dependence of $\Gamma$ on $g_v$ cannot be determined by the
bosonization approach\cite{doraphonon}.
The $\Gamma<\Gamma_1\approx \Gamma_2(1-\Gamma_2^2/\omega_0^2)$ region corresponds to underdamped phonons, while the
overdamped response shows up at $\Gamma_1<\Gamma<\Gamma_2$ with two distinct dampings.

\textit{Conductances at low and high temperatures.} The electric ($G$) and heat conductance ($\kappa$) through the molecular
transport junction in the
wideband limit are given by\cite{giuliano}
\begin{gather}
\left[\begin{array}{c}
G\\
\kappa
\end{array}\right]=
\frac{\Delta}{h}\int \textmd{d}\omega \frac{\partial f}{\partial\omega}
\left[\begin{array}{c}
e^2\\
T\omega^2
\end{array}\right]\textmd{Im}G_d(\omega),
\label{elcond}
\end{gather}
where $f$ is the Fermi function, and $G_d(\omega)$ is Green's function of the electron on the molecule.
From Eq. \eqref{trafo},
\begin{equation}
G_d(\omega)=\bar\nu^2 G_\Psi(\omega)
\label{ftopsi}
\end{equation}
at $x=0$. This basic relation allows us to determine the properties of localized electron from the Yu-Anderson model.
At $T=0$, one can derive a Fermi liquid relation for the Green's function of the $\Psi$ field\cite{cornaglia1,doraelastic} at $x=0$ 
as
\begin{equation}
G_\Psi(\omega=0)=-i\pi\rho_0,
\end{equation}
which holds true even in the presence of vibrational modes, i.e. the zero temperature zero frequency density of states remains
unchanged. This occurs because at $T=0$, the incoming electrons experience a frozen Fermi see and no bosons in the oscillator, thus
no phase space for scattering.
Identical results are obtained for the Kondo model.
This leads to the conductance at $T=0$ as
\begin{gather}
G(T=0)=\frac{e^2}{h}\frac{\Delta^2}{\Delta^2+\epsilon_0^2},
\label{g0}
\end{gather}
which ranges from perfect transmission ($G(0)=e^2/h$) for $|\epsilon_0|\ll\Delta$ to a decent suppression of the
conductance to $\sim 0.7-0.8 e^2/h$ for $|\epsilon_0|\lesssim \Delta$.
Eq. \eqref{g0} is expected to hold for low transparency junctions, beyond the validity of our mapping as well.
The heat conductance satisfies
\begin{equation}
\lim_{T\rightarrow 0}\frac{\kappa(T)}{T}=L_0 G(T=0),
\label{kappa0}
\end{equation}
where $L_0=(\pi k_B/e)^2/3$ is the Lorentz number, thus the Wiedemann-Franz law is fulfilled.
At high temperatures ($T>$max$(|\omega_{p\pm}|,|\omega_{p+}\omega_{p-}|/\Gamma$), but still obeying to $T\ll\Delta$,  the
$\mathcal{T}$-matrix
for the
$\Psi$ field reaches its universal value,
$\mathcal{T}=1/i2\pi\rho_0$.
This halves the
corresponding
Green's function at $x=0$ in this high $T$ region as
\begin{equation}
G_\Psi(\omega)=G_\Psi^0+
G_\Psi^0\mathcal{T} G_\Psi^0=-\frac{i\pi\rho_0}{2}
\end{equation}
with $G_\Psi^0=-i\pi\rho_0$, which determines the conductance at high temperatures as
\begin{equation}
G(T\gg\omega_0)=\frac{G(T=0)}{2}.
\end{equation}
The phonon state is populated by many bosons at high temperatures, and every incoming electron scatters off
them inelastically with increasing probability even in the weak coupling limit.
The inelastic scattering cross section reaches its maximal value\cite{doraelastic} at
this temperature range for arbitrary electron-vibration coupling.
The electrons  dephase completely, and can scatter forward and backward with equal probability (=1/2).
Consequently, the $\mathcal{T}$-matrix takes the above universal value, and the
conductance halves.
Similar conclusions about halving the spectral weight were reached for scattering on a dynamical boundary condition\cite{fuentes}.

\begin{figure}[h!]
\centering{\includegraphics[width=6cm,height=6cm]{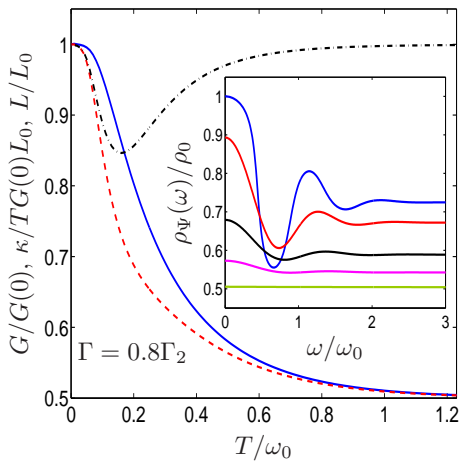}}
\caption{(Color online) The electric and heat conductance (blue solid and red dashed line) through a molecular junction and their
ratio (black dashed-dotted line)
is shown as a function of the temperature for $\Gamma=0.8\Gamma_2$, $W=10\omega_0\Rightarrow \omega_{\pm}/\omega_0=\pm 0.44-i0.06$.
The high
($T\gg$Re$\omega_{p\pm}$) and low
($T\ll$Re$\omega_{p\pm}$) temperature parts are universal.
The inset shows the evolution of the spectral
function, $\rho_\Psi(\omega)=-$Im$G_\Psi(\omega)/\pi$  for $T/\omega_0=0$, 0.2, 0.4, 0.6 and 1.2 from top to bottom.
\label{lorentzg80w10}}
\end{figure}


In the above high $T$ limit, the heat conductance also satisfies $\kappa(T\gg\omega_0)/T=\frac{\pi^2}{3 e^2}G(T\gg \omega_0)$,
and the Wiedemann-Franz law holds.
Therefore, both the electric and heat conductances are expected to exhibit a step from their $T=0$ value to half of its value with
increasing temperatures, and the Wiedemann-Franz law  holds in the two limits. In between the two limits, $G$ and $\kappa$ follows
a different temperature dependence: the heat conductance drops faster with the temperature, and their ratio stays below the
universal Lorentz number, hence the Wiedemann-Franz law is violated, as is shown in Fig. \ref{lorentzg80w10}.
The minimum of their ratio occurs roughly at $3T=$Re$\omega_{p\pm}$, which
can be used as a rule of thumb to estimate the
value of the renormalized phonon frequency.
We mention, that by further increasing the temperature to leave the $T\ll\Delta$ regime,
Eq. \eqref{hmol} practically decouples from the conduction electrons. In this regime, the conductances
exhibit a second step to zero.

\begin{figure}[h!]
\includegraphics[width=2.9cm,height=3cm]{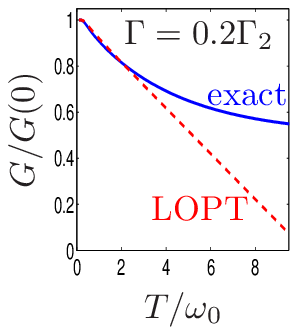}
\centering{\includegraphics[width=5.4cm,height=3cm]{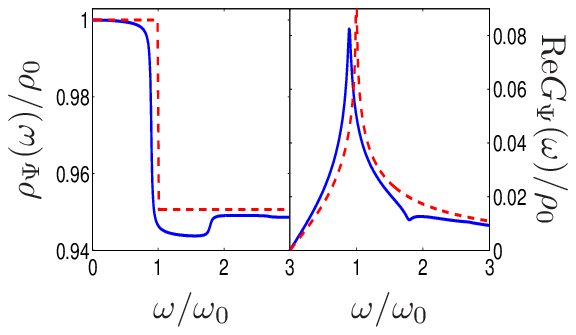}}
\caption{(Color online) The electric conductance from bosonization (blue solid) and from LOPT (red
dashed line) is shown in the left panel as a function of the temperature for weak electron-vibration coupling $\Gamma=0.2\Gamma_2$,
$W=10\omega_0 \Rightarrow \omega_{\pm}/\omega_0=\pm 0.89-i0.02$,
signaling the limitations of the LOPT.
The deviations grow with the coupling.
The right panel shows the real and imaginary part of the Green's function together with the LOPT result at $T=0$.
\label{condpert}}
\end{figure}

\textit{Green's function from bosonization.}
To study the crossover between the high and zero temperature limits, we use the exact result for
the Yu-Anderson
model obtained
from bosonization\cite{doraphonon,doraelastic} valid for arbitrary temperatures, whose derivation is sketched below.
From Eq. \eqref{ftopsi}, the electron Green's function on the molecule is related to that of the $\Psi$ field.
The real time dependence of the latter at the impurity site is obtained in the Yu-Anderson model as 
\begin{gather}
G_{\Psi}(t)=-i\frac{\Theta(t)}{4}\sum_{\gamma,\gamma'=\pm}\lim_{x,y\rightarrow 0^+}
\langle\{\Psi(\gamma x,t),\Psi^+(\gamma'y,0)\}\rangle=
\nonumber\\
=-i\frac{\Theta(t)}{8\pi}\sum_{\gamma,\gamma'=\pm}\left(\exp[C_{\gamma,\gamma'}(t)]+
\exp[C_{\gamma',\gamma}(-t)]\right),
\label{green1}
\end{gather}
where $C_{\gamma,\gamma'}(t)=\lim_{x,y \rightarrow 0^+}4\pi\langle\Phi(\gamma
x,t)\Phi(\gamma'y,0)-[\Phi(\gamma x,t)^2+\Phi(\gamma'y,0)^2]/2\rangle-
\ln(\alpha)$, and $\alpha\sim 1/W$ is the short distance cutoff in the bosonized theory, $\gamma$ and
$\gamma'$ denotes the sign of the spatial coordinates $x$ and $y$.
The expectation value $C_{\gamma,\gamma'}(t)$ at
bosonic Matsubara frequencies can be evaluated from the bosonized
Hamiltonian, Eq. \eqref{hamboson} as
\begin{gather}
C_{\gamma,\gamma'}(\omega_m)=-\frac{1}{4|\omega_m|}+\frac{\Gamma}{2}\frac{(\textmd{sgn}(\omega_m)+\gamma)
(\textmd{sgn}(\omega_m)-\gamma')}{(|\omega_m|+i\omega_{p+})(|\omega_m|+i\omega_{p-})},
\label{Gphi}
\end{gather}
where $\omega_m=2m\pi T$.
The first term is responsible for the $1/t$ decay of the local fermionic propagator, while the second one stems from the
interaction of electrons with the oscillator. For $\gamma=\gamma'$, this correction term vanishes.
$C_{\gamma,\gamma'}(t)$ is evaluated by analytically continuing the Matsubara frequencies to real ones in Eq. \eqref{Gphi}, which
defines the
spectral intensity, and then following Ref. \onlinecite{zubarev}, we arrive to the desired correlator.
By defining the integral,
\begin{gather}
A(t)=\int\limits_{-\infty}^{\infty}d\omega \frac{-i\exp(-i\omega
t)}{1-\exp(-\omega/T)}\frac{4\Gamma}{(\omega-\omega_{p+})(\omega-\omega_{p-})},
\end{gather}
which follows the derivation of the position autocorrelator of a harmonic oscillator coupled to a heat
bath\cite{weiss}, and can be evaluated in closed form using the hypergeometric functions, the local retarded
Green's function at finite temperatures follows as
\begin{gather}
G_{\Psi}(\omega,T)=-i\pi\frac{\rho_0}{2}\left(1+I_{ph}(\omega,T)+\exp\left({A(0)}\right)\right),
\label{retgreenT}
\end{gather}
where
\begin{gather}
I_{ph}(\omega,T)=\int\limits_0^\infty {dt} \frac{T\exp(i\omega t)}{\sinh(\pi T t)}\textmd{Im}\exp\left(
{A(t)}\right),
\end{gather}
whose $T=0$ limit was analyzed in Ref. \onlinecite{doraelastic}, giving Im$G_{\Psi}=-i\pi\rho_0$ at $T=\omega=0$.
In the $T\gg |\omega_{p+}\omega_{p-}|/\Gamma$ limit, $I_{ph}$ vanishes, and the imaginary part of the propagator is
Im$G_{\Psi}=-i\pi\rho_0/2$, which
is the desired result.
Plugging Eq. \eqref{retgreenT} to \eqref{ftopsi}, the conductances are evaluated from
Eq. \eqref{elcond}, which
are shown in Fig. \ref{lorentzg80w10} for the full $T$ range, which agree nicely with the analysis of
limiting cases.
The inset shows the local spectral function, the Franck-Condon steps\cite{flensberg,koch} arising from multivibron excitations are 
smooth due to the significant phonon
damping. Notice its explicit temperature 
dependence through the phonon occupation
number.
The power of the exact solution manifests itself in comparison with lowest order perturbation theory
(LOPT)\cite{engelsberg,eggergog}, which
neglects multiphonon contributions (becoming dominant with temperature), frequency renormalization and lifetime effects, as is
visualized in Fig. \ref{condpert}. LOPT
predicts
\begin{equation}
\frac{\rho_{\Psi}(\omega)}{\rho_0}=1-\frac{\Gamma\pi}{\omega_0}\left[\coth\left(\frac{\omega_0}{2T}\right)+
\sum_{s=\pm}sf(\omega+s\omega_0)\right],
\end{equation}
which breaks down completely at $T\gg (\omega_0,\omega_0^2/\Gamma)$ even in the
weak-coupling regime, leading to the complete
suppression  and even a sign change in $G$ at high temperatures.

\textit{Discussion.}
Experimentally, the renormalized vibration frequency is determined from the current-voltage characteristics: the $dI/dV$ exhibits
a step down in the large contact broadening limit\cite{eggergog} at $eV=$Re$\omega_{p\pm}$, which sets the characteristic temperature 
range of the predicted conductance change.
As a very rough estimate, the $T=0$ conductance is $\sim \rho_f(\omega\rightarrow eV)$, showing the multiphonon structures with the 
Franck Condon suppression\cite{flensberg} in the 
inset of 
Figs.~\ref{lorentzg80w10} and \ref{condpert}.
We mention, that low transmission junctions exhibit a step up in both the out-of-equilibrium $dI/dV$ and the equilibrium spectral
function at the excitation of the first vibrational mode.
Temperature dependent transport is feasible on molecular devices\cite{parks}, albeit large temperature variations are often
accompanied by structural changes or the mechanical deformation of the junction.
Therefore, the basic ingredients for the observation of
the predicted conductance step are the extreme mechanical stability of the device and a low enough frequency of the vibrational mode.
Fullerenes like C$_{60}$ were found\cite{park} to possess a center-of-mass oscillation of 50~K, which can be lowered by
considering heavier members of their family (e.g. C$_{140}$).
In this case, the oscillation frequency can accurately be estimated based on the interactions (electrostatic, van der Waals) between
the molecule and the electrodes.
In addition, the intercage vibrational modes of C$_{140}$ start from 25~K\cite{pasupathy}.
Another promising configuration, using a suspended quantum dot phonon cavity\cite{weig}, possesses a vibrational frequency of 0.8~K.

Although the electron-vibration coupling ($g_v$) is hardly controllable, the parameters $\Delta$ and $\rho_0$ can be
tuned by varying the contact DOS, which
can drive the system towards stronger effective couplings. This decreases the temperature window for the conductance step.
Given a low vibrational frequency, condition $\Delta\gg (T,\omega_0)$ is easily met, therefore we expect that a dedicated setup
allows the
observation of the universal conductance steps, while low temperature $dI/dV$ measurements can reveal the multiphonon 
structures in the local spectral function on the molecule. Even for molecular devices, where the direct temperature 
dependence cannot be traced, our results demonstrate the role of multiphonon scattering processes in the room temperature 
conductance of the junction.

In summary, we studied non-perturbatively the electric and heat conductance through a single level coupled to a vibrational mode by 
mapping it to the Yu-Anderson model.
With increasing temperature, the conductances drop to half of their $T=0$ value due to increasing inelastic scattering
 in the large contact 
broadening limit.
We argue that its experimental observation is within reach in stable contacts with a low vibrational mode.

\begin{acknowledgments}
This work was supported by the Hungarian
Scientific Research Funds under grant numbers K72613 and K76010, and by the Bolyai program of the Hungarian Academy of Sciences.
\end{acknowledgments}

\vspace*{-8mm}

\bibliographystyle{apsrev}
\bibliography{wboson}
\end{document}